\documentclass{article}
\usepackage{spconf,amsmath,graphicx}
\usepackage{subfigure}
\usepackage{booktabs} 

\usepackage{tikz}
\usepackage{amsmath}
\usepackage{caption}
\usepackage{adjustbox}
\usepackage{multirow}
\usepackage{multicol}
\usepackage{enumitem}
\makeatletter
\def\BState{\State\hskip-\ALG@thistlm}
\makeatother
\title{Workload-aware Automatic Parallelization \\for Multi-GPU DNN Training}
\name{\begin{tabular}{c}Sungho Shin$^{\dagger}$, Youngmin Jo$^{\dagger}$, Jungwook Choi$^{\star}$,\\
Swagath Venkataramani$^{\star}$, Vijayalakshmi Srinivasan$^{\star}$, and Wonyong Sung$^{\dagger}$\thanks{This work was supported in part by the Brain Korea 21 Plus Project and the National Research Foundation of Korea (NRF) grant funded by the Korea government (MSIP) (No. 2018R1A2A1A05079504).}\end{tabular}}
\address{$^{\dagger}$Department of Electrical and Computer Engineering, Seoul National University, Seoul, 08826 Korea\\$^{\star}$IBM Research AI, 1101 Kitchawan Rd. Yorktown Heights, New York 10598\\
sungho.develop@gmail.com, youngmin.research@gmail.com, choij@us.ibm.com,\\ swagath.venkataramani@ibm.com, viji@us.ibm.com, wysung@snu.ac.kr}

\begin{document}
\ninept
\setlength{\textfloatsep}{2pt}

\maketitle
\begin{abstract}
Deep neural networks (DNNs) have emerged as successful solutions for variety of artificial intelligence applications, but their very large and deep models impose high computational requirements during training. Multi-GPU parallelization is a popular option to accelerate demanding computations in DNN training, but most state-of-the-art multi-GPU deep learning frameworks not only require users to have an in-depth understanding of the implementation of the frameworks themselves, but also apply parallelization in a straight-forward way without optimizing GPU utilization. In this work, we propose a workload-aware auto-parallelization framework (WAP) for DNN training, where the work is automatically distributed to multiple GPUs based on the workload characteristics. We evaluate WAP using TensorFlow with popular DNN benchmarks (AlexNet and VGG-16), and show competitive training throughput compared with the state-of-the-art frameworks, and also demonstrate that WAP automatically optimizes GPU assignment based on the workload's compute requirements, thereby improving energy efficiency. 
\end{abstract}
\begin{keywords}
Multi-GPU training, data parallelization, auto parallelization, neural network training, deep learning framework
\end{keywords}
\section{Introduction}
\label{sec:introduction}
In recent years, deep learning (DL) has emerged as the dominant solution showing remarkable success in a wide spectrum of artificial intelligence (AI) applications~\cite{krizhevsky2012imagenet, simonyan2014very, he2016deep, graves2006connectionist, graves2013speech, cho2014learning, shin2016dynamic}. 
In each of these domains, deep neural networks (DNNs) achieve superior accuracy through the use of very large and deep models -- necessitating up to 100s of ExaOps of computation during training. 

To deal with the high computational demands and to achieve high throughput, DNN training is often accelerated by parallelizing across multiple GPUs. Popular DL frameworks such as TensorFlow~\cite{TensorFlow2015-whitepaper} and Pytorch~\cite{pytorch2017} provide native GPU support. However, adapting a single-GPU DNN model to work with multi-GPU environments is not trivial for users, since they must consider not only how computation will be distributed across multiple GPUs but also what data will be exchanged via communication between GPUs. A sub-optimal implementation decision can easily lead to poor GPU utilization causing a significant drop from the expected speedup due to parallelization.

There have been several TensorFlow based parallelization frameworks to ease the burden of multi-GPU implementation for the users, such as Parallax~\cite{kim2018parallax}. Parallax provides a set of Python-level APIs for the users to adapt their single GPU code for the multi-GPU runs. These APIs help specify detailed information to the DL framework about the DNN description as well as the list of available GPUs. 
Using such information Parallax distributes the computations across GPUs and executes them in parallel. These frameworks also adopt the popular communication protocols such as Open MPI~\cite{gropp1999using} and NVIDIA collective communications library (NCCL)~\cite{nccl} for efficient data communication. However, they do not take into account GPU utilization during parallelization. Instead, these frameworks distribute the workload to all the available GPUs oblivious to the users. It is well known that GPU suffers low utilization when the workload is not sufficiently large, e.g., when minibatch size is small in DNN training. Thus, it has been so far the users' responsibility to determine the optimal number of GPUs based on the DNN workloads, and accordingly utilize the APIs. 

In this work, we propose a novel workload-aware automatic parallelization framework (WAP). Different from the existing frameworks, WAP applies parallelization under the hood of TensorFlow source code. At this stage the computation workloads across the DNN layers are fully specified into a dataflow graph and ready for analysis. 
We devise the workload analysis unit in our parallelization framework to 
estimate the expected utilization of the GPUs and determine the best number of GPUs to be assigned. Based on this analysis, we directly modify TensorFlow's dataflow graph to seamlessly enable multi-GPU parallelization. 
We automate all of these steps in WAP without requiring any additional inputs from the users. In particular, the users do not need to manually find out the number of GPUs optimally suited for a given workload.  We evaluate WAP with the popular DNN benchmarks, and demonstrate that it not only achieves competitive throughput and scalability compared to the state-of-the-art multi-GPU framework~\cite{kim2018parallax}, but also automatically optimize the GPU assignment based on the workload analysis, improving energy efficiency.

\begin{figure}[t]
\centering
    \includegraphics[width=\linewidth]{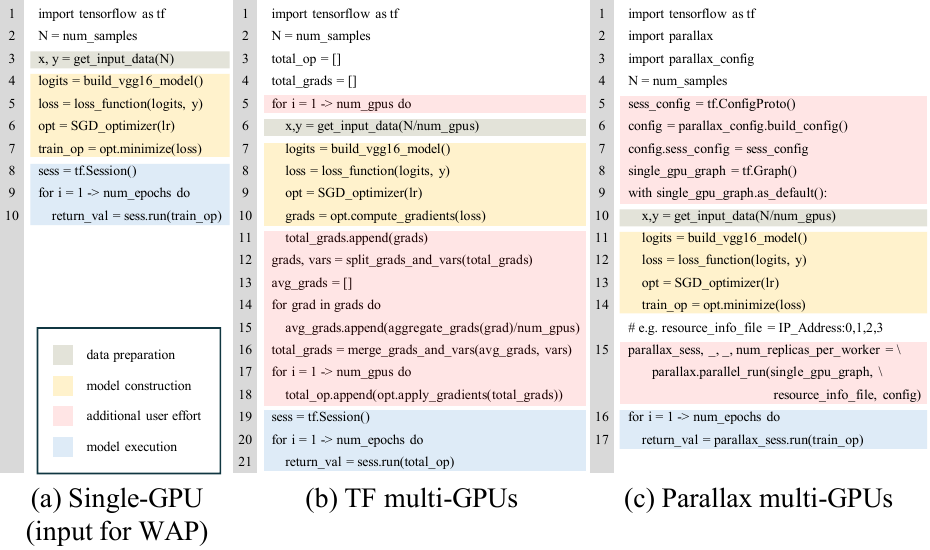}
\caption{Example pseudocodes for a single or multi-GPU training. (a), (b): single and multi-GPU implementation using default TensorFlow~\cite{TensorFlow2015-whitepaper}, and (c) multi-GPU with Parallax~\cite{kim2018parallax}. The red boxes indicate an additional user effort to execute multi-GPU training. Note that WAP can execute multi-GPU training only using (a). 
}
\label{fig:code}  
\end{figure}

\section{Related Work}
\label{sec:related_works}
\subsection{Parallelization Strategies for DNN Training}
\label{subsec:multi-GPU_training}
There have been extensive research on parallelization of DNN training~\cite{krizhevsky2012imagenet, dean2012large, krizhevsky2014one}. Most efforts can be categorized into three strategies: data, model, and hybrid parallelization. In data parallelization, each GPU uses the same DNN model (i.e., ``Replicated-Variables'' in~\cite{pr}) to train on a different subset of training data and compute gradients, which need to be aggregated across the GPUs~\cite{krizhevsky2012imagenet}. In model parallelization, a DNN model is split and distributed across multiple GPUs and each GPU is responsible only for updating a portion of the model~\cite{dean2012large}. Hybrid parallelization blends model and data parallelization; e.g.,  \cite{krizhevsky2014one} employs data and model parallelization for convolution and fully-connected layers, respectively. In this work, we focus on data parallelization as it is one of the most popular multi-GPU training schemes. However our under-the-hood parallelization based on TensorFlow dataflow graph is not limited to data parallelization; implementation of model and hybrid parallelization strategies will be future work.



\subsection{Multi-GPU Data-Parallel Training Frameworks}
\label{subsec:frameworks}
Many deep learning frameworks provide native GPU support to exploit its extensive parallel computing power in accelerating DNN training~\cite{TensorFlow2015-whitepaper,pytorch2017,dynet}. However, most of them require non-trivial manual efforts for converting a single-GPU code into a multi-GPU version. As an example, \figurename~\ref{fig:code} shows how a single-GPU implementation of VGG-16 is converted in TensorFlow and Parallax for multi-GPU runs. As shown in \figurename~\ref{fig:code}(b), TensorFlow requires users to handle details of multi-GPU implementation, such as the replication of DNN models (line 5-11) and the gradient aggregation (line 12-18).

To ease the users' burden of multi-GPU implementation, several frameworks provide Python-level APIs, and Parallax~\cite{kim2018parallax} is one of the most recent development. Parallax provides a software API for efficient distributed training (using Horovod~\cite{sergeev2018horovod} for efficient data communication) that hides detailed parallelization settings from the users, as shown in \figurename~\ref{fig:code}(c). While Parallax alleviates user effort for multi-GPU DNN training, it does not take into account GPU utilization during parallelization. As shown in \figurename~\ref{fig:code}(c), Parallax simply uses the list of GPUs available to the user, regardless of the amount of work for each layer of the neural network (line 15). Therefore, even if the GPU utilization is low due to a small workload size for a given DNN layer (e.g., when minibatch size is small), Parallax obliviously allocates all the GPUs, potentially wasting power and degrading performance due to unnecessary communication overheads.

In this work, we set out to realize auto-parallelization while being cognizant of the expected GPU utilization.
Unlike other existing multi-GPU frameworks, our framework provides automatic parallelization starting from a single-GPU code from the user, 
and takes into consideration the workload's compute requirements to optimally choose the number of GPUs for parallelization so as to increase both throughput and energy efficiency.

\section{Workload-aware Automatic Parallelization}
\label{sec:awwp}

\subsection{Overview}
\label{subsec:overview}

In this section, we explain details of our workload-aware automatic parallelization (WAP) framework.  
The key challenge in WAP is to extract the total amount of computations for all the DNN layers, including minibatch size, before applying parallelization. In TensorFlow, a dataflow graph created inside the TensorFlow core source code captures all the workload details of a given DNN. Therefore, we decided to augment the TensorFlow core source code for workload-aware parallelization.
\begin{figure}[t]
\centering
    \includegraphics[width=.8\linewidth]{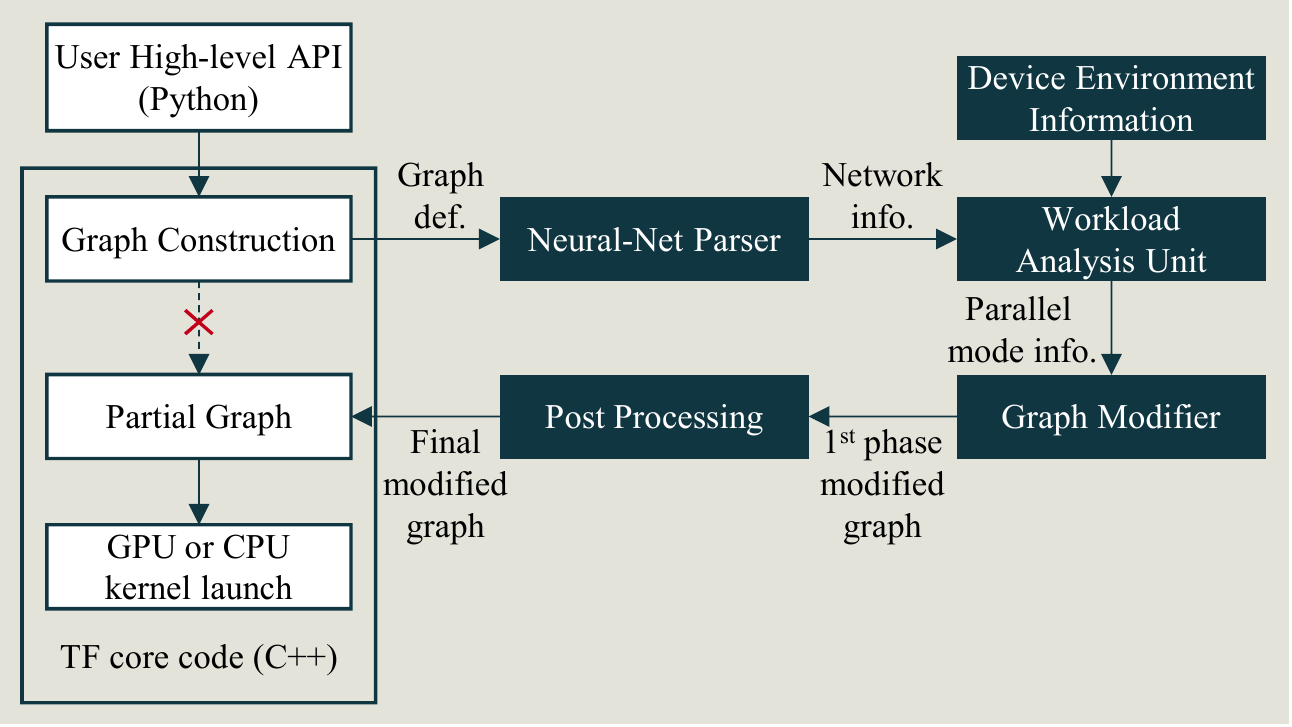}
\caption{Overview of workload-aware auto parallelization (WAP). The white boxes represent execution steps in TensorFlow core source code, and the navy boxes are added steps for WAP.}
\label{fig:overview}  
\end{figure}
\begin{figure}[t]
    \centering
    \hfill
    \subfigure[A dataflow graph from Single-GPU code]{\label{fig:1a}\includegraphics[width=0.205\textwidth,trim={0 1.3cm 0 0.3cm },clip]{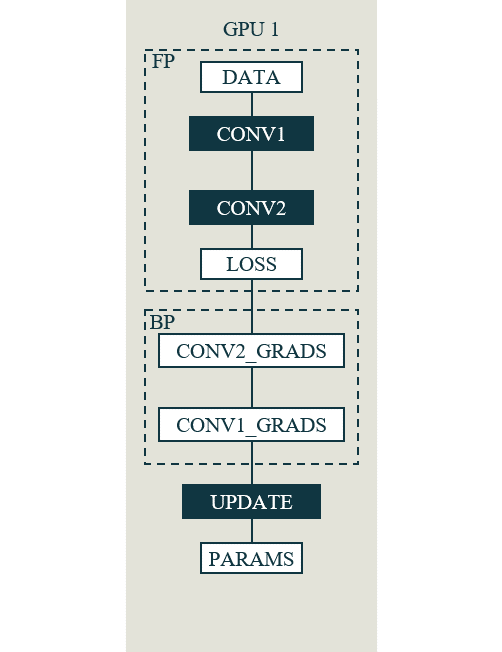}}
    \hfill
    \subfigure[Step1 (node replication)]{\label{fig:1b}\includegraphics[width=0.21\textwidth,trim={0 1.3cm 0 0.3cm },clip]{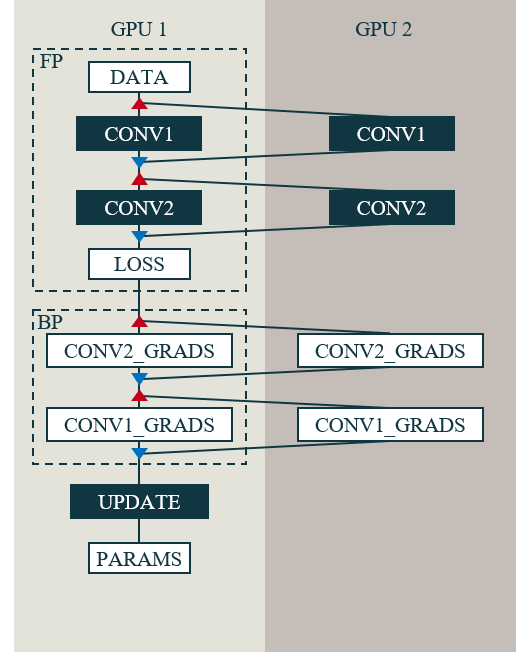}}\\
    \hfill
    \subfigure[Step2 (removing overhead)]{\label{fig:1c}\includegraphics[width=0.21\textwidth,trim={0 0.3cm 0 0.3cm },clip]{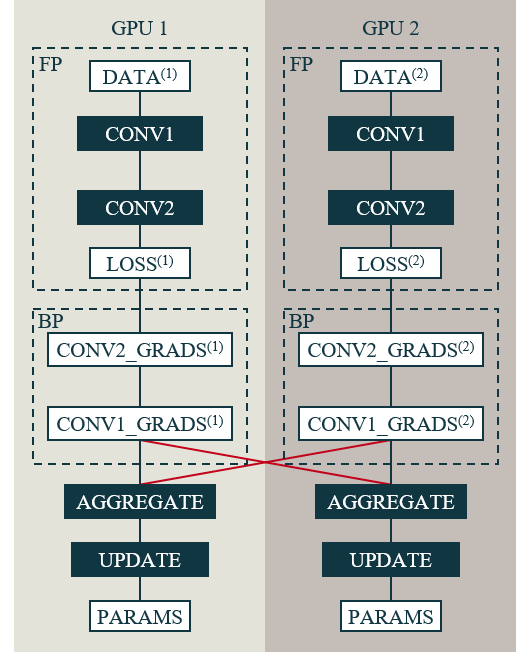}}
    \hfill
    \subfigure[Step3 (efficient gradient aggreation)]{\label{fig:1d}\includegraphics[width=0.21\textwidth,trim={0 0.3cm 0 0.3cm },clip]{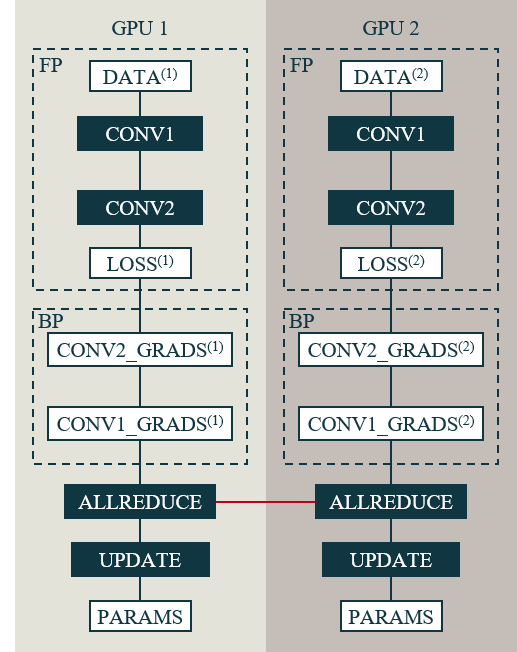}}

    \caption{Illustration of workload-aware parallelization (WAP). (a) A dataflow graph for Single-GPU code. (b) The primary computation nodes are replicated with ``split" and ``concatenate" nodes (red and blue triangles, respectively). (c) Unnecessary data communication between devices are removed. (d) Gradients aggregation is optimized with NCCL AllReduce.}
    \label{fig:1}
\end{figure}

~\figurename~\ref{fig:overview} shows the overall steps of WAP. The dataflow graph constructed from the Python-level user description of a DNN is sent to a \textit{Neural-Net Parser} to extract workload information. This information is used by the \textit{Workload-aware Analysis Unit} (WAU) to determine the optimal number of GPUs for parallelization. Next, the original dataflow graph is transformed via \textit{Graph Modifier} to reflect the GPU parallelization choice from WAU, followed by \textit{Post Processing} of the graph to avoid any redundancy in data communication. The final modified graph is then registered as the new graph object into TensorFlow, where each GPU is allocated a partial graph for parallel execution.


There are two advantages from our approach: 1) we extract the workload information at the same level as TensorFlow core does, so that we can perform precise workload analysis for the best GPU usage, and 2) by applying parallelization at the dataflow graph level under the hood, we hide all the burden of multi-GPU parallelization from the users. 
In the following subsection, we explain the important modules of WAP shown in~\figurename~\ref{fig:overview}.

\subsection{Implementation Details}
\label{subsec:implementation}


\subsubsection{Workload Analysis Unit}
\label{subsubsec:workload}
Workload analysis unit (WAU) receives the device (i.e., available GPUs) and the neural network information to determine how to perform data parallelization with the highest GPU utilization. The more GPUs are used, the more fine-grain the workload is divided, at some point making the amount of computation not enough to maintain high GPU utilization. At the same time, using more GPUs increases the communication overhead for the gradient aggregation. Therefore, sometimes using more GPUs achieves slower training speed. WAU analyzes the workload using the performance model to detect and use only the number of GPUs required to improve both throughput and energy efficiency.

There are several choices available to estimate expected performance. In this work, we adopt the GPU execution time model from ~\cite{zhihao2018} to evaluate run-time performance as follows: 
\begin{align}
	t_{\mathrm{estimate}} = \sum_{l_i}{\{t_c(l_i,d) + t_s(l_i,d)\}} \label{eq:tdgkt}
\end{align}
where $t_{\mathrm{estimate}}$ is the estimated total execution time for the entire layers of a DNN, $l_i$ denotes $i$-th layer in the network, $d$ is a factor for workload division, $t_c$ is the processing time for the forward and backward propagation, and $t_s$ represents the data communication time for gradient aggregation. We change $d$ from 1 to the total available GPUs to find out how many GPUs are needed to minimize $t_{\mathrm{estimate}}$. In Section~\ref{subsec:workload} we show that this performance analysis successfully guides to find the right parallelization that maximizes throughput and avoids unnecessary power consumption.

\subsubsection{Neural-Net Parser and Graph Modifier}
\label{subsubsec:nodes}

\textit{Neural-Net Parser} identifies which computational workloads need to be parallelized based on the parallelization strategies. In this work, it finds the computationally challenging layers such as convolution, and fully-connected layers from the dataflow graph and extract the workload information for analysis in WAU. 

\textit{Graph Modifier} transforms the original dataflow graph into the multi-GPU version as shown in~\figurename~\ref{fig:1}. Based on the number of GPUs determined by WAU, it first replicates the primary computation nodes (e.g., convolution and fully-connected), and split/concatenate their input and output, respectively (\figurename~\ref{fig:1b}). Note that naive node replication in this step causes unnecessary links for data communication across GPUs. In order to avoid such communication overhead, \textit{Graph Modifier} further replicates auxiliary computation nodes (e.g., activation functions, loss computation, etc.) and removes unnecessary split/concatenation (\figurename~\ref{fig:1c}). 


\subsubsection{Post Processing for Optimizing Gradient Reduction}
\label{subsubsec:comm_opt}
As explained in Section~\ref{subsec:multi-GPU_training}, in data parallelization, gradients calculated within each GPU need to be aggregated for the weight update. Naive implementation of this gradient aggregation can cause significant overhead, since it requires all-to-all data communication as shown in~\figurename~\ref{fig:1c}. The complexity of this communication overhead is O$({W}{N}^2)$, where the size of weight is $W$ and the number of GPUs is $N$. 
This communication overhead can be reduced by exploiting AllReduce with the ring algorithm, which reduces the complexity down to O$({W}{N})$~\cite{nccl}. As shown in \figurename~\ref{fig:1d}, in \textit{Post Processing}, we inserted NCCL AllReduce for efficient gradients aggregation.

\subsubsection{Evaluation}
\label{subsubsec:eval}
Each of the steps above in our graph-based data parallelization method is critical for achieving the best parallelization performance. To quantify the importance of each step, \tablename~\ref{table:after_eval} reports the impact of each step to the training throughput. Note that a naive replication of primary computation nodes in Step1 significantly degrades the throughput. Replicating auxiliary computation nodes accordingly with the removal of redundant data communication in Step2 helps recover the throughput. Optimization via NCCL AllReduce in Step3 further increases the throughput by $9\%$.

\begin{table}[t]
\caption{Training throughput (images/sec) of the WAP graph transformation steps on AlexNet~\cite{krizhevsky2012imagenet} with four GPUs and 2048 minibatch. Note that ``before'' represents single GPU performance. Experiments are conducted on ``single machine" which is explained in Section~\ref{subsec:setup}.}
\label{table:after_eval}
\centering
\begin{tabular}{ccccc}
\hline\hline
        & Before & Step1 & Step2 & Step3 \\ \hline
AlexNet & 2482     & 421   & 7264  & 7904  \\ \hline\hline
\end{tabular}
\end{table}

\begin{figure*}[t]
    \centering
    \subfigure[AlexNet-SM]{\label{fig:3a}\includegraphics[width=0.41\linewidth]{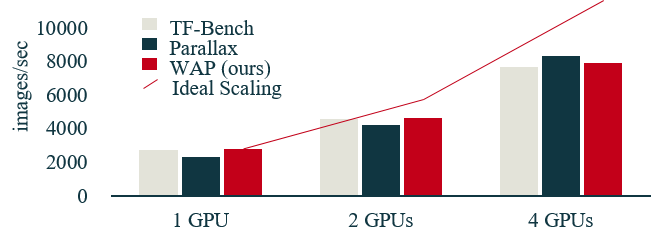}}
    \subfigure[AlexNet-DGX]{\label{fig:3b}\includegraphics[width=0.41\linewidth]{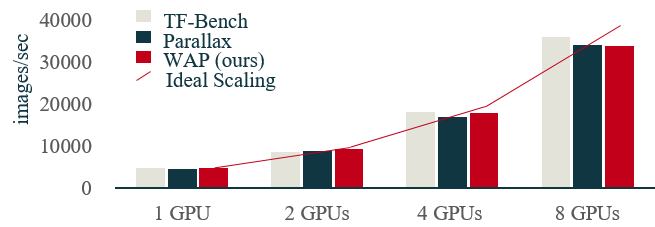}}
    \\
    \subfigure[VGG-16-SM]{\label{fig:3c}\includegraphics[width=0.41\linewidth]{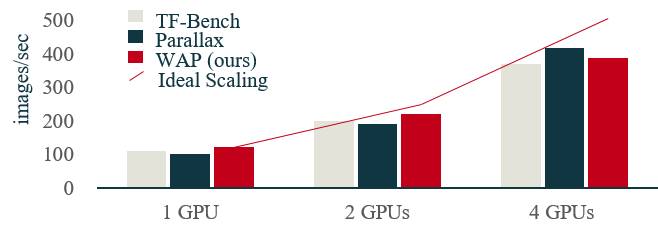}}
    \subfigure[VGG-16-DGX]{\label{fig:3d}\includegraphics[width=0.41\linewidth]{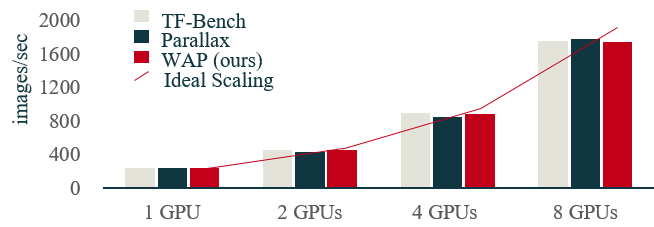}}

    \caption{Comparison of training throughput (images/sec) for AlexNet and VGG-16 on Single Machine (SM) and NVIDIA DGX-1 (DGX). WAP demonstrates compelling performance and favorable scalability without asking user effort for multi-GPU runs}
    \label{fig:3}
\end{figure*}

\section{Experimental Results}
\label{sec:experimental}
\subsection{Experimental Setup}
\label{subsec:setup}
All experiments are performed on two environments: A single machine with 4 GPUs (SM) and NVIDIA DGX-1 (DGX)~\cite{dgx}. SM is equipped one AMD Ryzen Threadripper 1900X 8-core CPUs, 96GB main memory, and four NVIDIA TitanXP GPUs (connected via PCIe). DGX is equipped with dual 20-core Intel Xeon ES-2698, 512GB main memory, and eight NVIDIA Tesla GP100 GPUs (connected via NVLink). We evaluate our framework with the two well-known convolutional neural networks (CNNs), AlexNet~\cite{krizhevsky2012imagenet} and VGG-16~\cite{simonyan2014very}. We employ data-parallel training with 512 and 64 per-GPU minibatch for AlexNet and VGG-16, respectively. 

We compare the training throughput of WAP against two state-of-the-art multi-GPU implementations: TensorFlow high performance benchmark (TF-Bench)~\cite{tfbm} and Parallax~\cite{kim2018parallax}. Note that TF-Bench is manually coded with the parallelization details hand-optimized, and Parallax is coded with the provided API. Whereas, WAP does not need any change from the single-GPU code. For fair comparison, all three implementations employ the same optimization schemes, such as Replicated-Variables and AllReduce (TF-Bench and WAP use NCCL AllReduce, and Parallax uses AllReduce from Horovod), for the gradient aggregation. 


\subsection{Training Performance}
\label{subsec:AlexNet}
First, we evaluate the training performance of WAP in terms of scalability in throughput. 
\figurename~\ref{fig:3a} and \figurename~\ref{fig:3b} show the training throughput of AlexNet on SM and DGX, respectively. Similarly, \figurename~\ref{fig:3c} and \figurename~\ref{fig:3d} are for VGG-16 on SM and DGX, respectively. Throughout the experiments, WAP consistently demonstrates competitive performance. In particular, its throughput is in par with TF-Bench for the most cases, validating that the auto-parallelized execution of WAP is as good as the hand-optimized TF-Bench code. Note that Parallax shows slightly lower/higher performance with smaller/larger number of GPUs, respectively. This is in part due to the AllReduce implementation in Parallax; it employs Horovod's AllReduce, which reports better AllReduce performance with large number of GPUs~\cite{sergeev2018horovod}, but its MPI runs would suffer higher overhead when the number of GPUs is small. The results on DGX show better scalability than SM, since the communication overhead of DGX is further reduced thanks to NVLink. Overall, WAP achieves compelling performance and scalability without requiring manual user effort for multi-GPU runs.



\subsection{Workload-Aware GPU Allocation}
\label{subsec:workload}
We now showcase a scenario when the workload-aware GPU allocation ends up achieving higher speedup as well as saving power consumption. \tablename~\ref{table:estimate} shows the measured throughput and power consumption in the SM machine (with 4 GPUs) for training AlexNet with minibatch of 128. 

In case of Parallax, all four GPUs are obliviously used for data parallelization. Since each GPU gets 32-minibatch amount of workload, which is not large enough to achieve high GPU utilization, the speedup by parallelization is overshadowed by the increased data communication overhead. Thus Parallax suffers lower throughput using 4 GPUs than what it could achieve with 1 GPU. In case of WAP, however, the workload is first analyzed by WAU, where the estimated throughput from Equation~\eqref{eq:tdgkt} indicates that the 1-GPU run would outperform the 4-GPU run. Based on this analysis, WAP uses only 1 GPU and achieve higher throughput. This demonstrates that WAU effectively hides the burden of optimizing GPU utilization from the users.

The workload-aware GPU allocation also has significant impact on energy efficiency. In case of Parallax, 4 GPUs are used (although each of them are running with lower utilization), thus it suffers high power consumption. In contrast, WAP only uses one GPU, reducing power consumption by $63\%$ compared to Parallax.


\begin{table}[t]
\centering
\caption{Comparison of throughput (images/sec) and power consumption (Watt) of Parallax and WAP on Alexnet with minibatch size of 128 on Single Machine (SM) with four GPUs. The numbers in parentheses mean ``Used GPUs''. Parallax obliviously uses four GPUs, while WAP chooses to use one GPU based on the workload estimation by WAU.}
\label{table:estimate}
\begin{tabular}{ccccc}
\hline\hline
\multirow{2}{*}{}                                                                  & \multirow{3}{*}{\begin{tabular}[c]{@{}c@{}}Available\\ GPUs\end{tabular}} & Parallax                                                         & \multicolumn{2}{c}{\begin{tabular}[c]{@{}c@{}}WAP (Ours)\end{tabular}}                                                       \\ \cline{3-5} 
                                                                                   &                                                                           & \begin{tabular}[c]{@{}c@{}}Measured \\ (used GPUs)\end{tabular} & \begin{tabular}[c]{@{}c@{}}Estimated \\ by  WAU\end{tabular} & \begin{tabular}[c]{@{}c@{}}Measured \\ (used GPUs)\end{tabular} \\ \hline
\multirow{2}{*}{\begin{tabular}[c]{@{}c@{}}Throughput\\ (images/sec)\end{tabular}} & 1                                                                         & 1986 (1)                                                         & 2244                                                         & 2560 (1)                                                         \\ 
                                                                                   & 4                                                                         & \textbf{1473 (4)}                                                         & 1491                                                         & \textbf{2560 (1)}                                                         \\ \hline
\multicolumn{2}{c}{Power (Watt)}                                                                                                                                & 402.81                                                           & \multicolumn{2}{c}{149.44}                                                                                                     \\ \hline\hline
\end{tabular}
\end{table}

\section{Concluding Remarks}
\label{sec:concluding}
In this work, we proposed a workload-aware automatic parallelization (WAP) framework for DNN training, which automatically distributes work to multi-GPUs based on the workload characteristics. 
The proposed tool is implemented on the TensorFlow core source code for executing multi-GPU training without any end-user's effort. WAP automatically modifies the single-GPU to multi-GPU graph with the significant consideration of communication cost and distribution of computational nodes. We evaluate WAP with popular DNN benchmarks (AlexNet and VGG-16), and show competitive training throughput compared with the state-of-the-art hand-optimized parallelization frameworks, and also demonstrate that WAP automatically optimizes GPU assignment based on the workload's compute requirements, thereby decreasing power consumption and improving overall energy efficiency.

\vfill\pagebreak
\bibliographystyle{IEEEbib}
\bibliography{refs}
\end{document}